**(Co, Ni)Sn$_{0.5}$ nanoparticles supported on hierarchical CN-graphene-based electrocatalysts for the oxygen reduction reaction**

Enrico Negro, Angeloclaudio Nale, Keti Vezzù, Federico Bertasi, Gioele Pagot, Yannick Herve Bang, Stefano Polizzi, Massimo Colombo, Mirko Prato, Laura Crociani, Francesco Bonaccorso, Vito Di Noto*

The synthesis of new *"Pt-free"* electrocatalysts (ECs) for the oxygen reduction reaction (ORR) is reported. The ECs are characterized by a hierarchical *"core-shell"* morphology; the *"core"* is made of graphene, that is covered by a cratered, microporous carbon nitride (CN) *"shell"*. The latter supports nanoparticles of M1 and Sn metals (M1 = Co; Ni) in *"coordination nests"*. These latter are holes in the CN matrix, whose walls consist of N- and C-ligands. Two groups of ECs are studied: (i) *"pristine"* ECs; and (ii) *"activated"* ECs, that are obtained from the *"pristine"* ECs by means of a suitable activation process (**A**) aimed at improving the performance in the ORR. Here is clarified the interplay existent between: (i) the chemical composition, morphology, structure and **A**; and (ii) the ORR performance and mechanism as a function of the pH of the environment. The resulting insights improve the fundamental understanding of this family of ECs and open the door to the devising of new preparations of *"Pt-free"* ECs for the ORR, which: (i) are stabilized by a CN matrix and; (ii) exhibit an improved performance.

**Keywords**

Graphene, Supported catalysts, Structure-activity relationships, Oxygen reduction reaction, CV-TF-RRDE method.

## 1. Introduction

The sluggishness of the oxygen reduction reaction (ORR) is still a crucial limiting factor towards the widespread implementation of a number of promising, efficient and environmentally-friendly electrochemical energy conversion and storage (EECS) devices such as metal-air batteries and fuel cells operating at T < 250°C (*e.g.*, proton-exchange membrane fuel cells, PEMFCs, high-temperature proton-exchange membrane fuel cells, HT-PEMFCs, phosphoric acid fuel cells, PAFCs, and anion-exchange membrane fuel cells, AEMFCs).[1] Indeed, the ORR plays an important role in the operation of the latter EECS devices. From a fundamental viewpoint, the poor kinetics of the ORR introduces large activation overpotentials, $\eta_{ORR}$, on the order of *ca.* 270-300 mV or more.[2] The latter erode the energy conversion efficiency of all the EECS devices relying on the ORR, reducing the competing advantage towards traditional technologies (*e.g.*, internal combustion engines, ICEs).[3] In practice, $\eta_{ORR}$ is minimized by using suitable electrocatalysts (ECs); unfortunately, at T < 250°C the ECs affording the lowest $\eta_{ORR}$ include a high loading of platinum-group metals (PGMs),[1a] which are characterized by a very low abundance in Earth's crust and raise large risks to incur in supply bottlenecks.[4] On these basis, the development of high-performing ECs for the ORR that do not require PGMs (typically indicated as *"Pt-free"* ECs) is a major goal of both fundamental and applied research.

A high-performing EC for the ORR must exhibit the following main features: (i) a minimized $\eta_{ORR}$, that is generally achieved by modulating the chemical composition and maximizing the surface density of the active sites; (ii) a high electrical conductivity, to curtail the ohmic losses; and (iii) a facile transport of reactants and products to and from the active sites, to abate concentration losses.[1b, 5] *"Pt-free"* ECs are typically based on a carbonaceous matrix doped with heteroatoms [1e] such as B,[6] N,[7] S,[8] P,[9] I,[10] or a combination thereof.[11] Such a system can be used as an ORR EC by itself;[6, 12] however, in most instances the matrix coordinates either: (i) metal atoms [13] (typically Fe [7, 14] or Co [15]); or (ii) nanoparticles (NPs) based on Earth-abundant metals and one or more of C, N, O, S, Se.[16] Other *"Pt-free"* ORR ECs are based on inorganic nanostructures with the same range of chemical compositions as the NPs described above in (ii), without the need of a carbonaceous matrix.[17] However, despite a massive research effort currently taking place worldwide, the exact interplay between the chemical composition and the ORR performance/mechanism of *"Pt-free"* ECs is not understood clearly. In particular, for any given EC it is often difficult to identify what is the *"real"* ORR active site, not to mention that the literature generally disregards the role played by the pH of the environment. This is a serious shortcoming, since: (i) from a fundamental viewpoint, the pH of the environment strongly affects the ORR mechanism;[18] and (ii) each of the intended applications of the EC may force the ORR to occur at a different pH value, that may easily span from 0-1 (*e.g.*, in the case of PEMFCs) [19] to 13-14 (*e.g.*, in AEMFCs).[20]

The overall energy conversion efficiency of EECs at high current densities is significantly affected by the charge and mass transport issues of the ORR EC.[21] The latter features are mostly modulated by the composition and morphology of the EC and, in particular, of its support.[21] Typical supports of *"Pt-free"* ECs for the ORR consist of mesoporous carbons exhibiting a large surface area, on the order of 100-1000 $m^2 \cdot g^{-1}$,[12a, 22] intended to maximize the exposure/accessibility of the active sites. On the other hand, these systems may suffer from significant mass transport issues due to the small average size of the pores and the consequent tortuous paths that reactants and products must follow to reach the active sites.[23] One interesting possibility to address these points is to use graphene as the support of an ORR EC. Indeed, graphene exhibits a number of very attractive features, including: (i) an outstanding electron mobility (up to 200000 $cm^2 \cdot V^{-1} \cdot sec^{-1}$),[24] that translates into an outstanding electron conductivity; and (ii) a very large surface area, up to a theoretical value of *ca.* 2600 $m^2 \cdot g^{-1}$.[25] In particular, owing to its 2D morphology, graphene is not expected to exhibit microporosity, with its surface easily accessible to the environment. Finally, there is some evidence that graphene is able to improve the ORR performance of an EC by influencing the electronic features of its active sites (*e.g.*, through electron transfer phenomena).[26] These latter are not completely understood, and are outside the scope of this work; thus, they will not be discussed. In any event, the implementation of graphene as an EC support is hindered by its high chemical inertia.[27] The literature reports several approaches to use graphene in the synthesis of *"Pt-free"* ECs, that include: (i) impregnation of graphene oxide (GO) precursor with N- and M- species, followed by thermal reduction;[28] and (ii) hydrothermal processes,[29] among many others.

A unique synthetic protocol was devised and optimized in our laboratory since more than 15 years, that is able to yield ECs for the ORR based on a carbon nitride (CN) matrix.[5, 30] The proposed protocol is extremely flexible and allows for the preparation of two generations of ECs exhibiting a well-controlled chemical composition and morphology. *"First-generation"* ECs consist in metal alloy NPs embedded in graphite-like carbon nitride matrices of larger NPs.[31] It was found that these *"first-generation"* ECs yield the best performance in the ORR when the N wt% in the sample is on the order of 5% or less. *"Second-generation"* ECs exhibit a *"core-shell"* morphology, where the *"core"* consists of conductive NPs, that are covered by a CN *"shell"* embedding/coordinating the active sites for the ORR in C and N-based *"coordination nests"*.[5, 18b, 21a] The flexibility of this preparation protocol allows for the preparation of a variety of *"second-generation"* ECs, including either: (i) a low loading of PGMs (L-PGM);[21a, 32] or (ii) completely *"Pt-free"*.[18b] The ECs obtained with this preparation protocol are obtained starting with a precursor based on a zeolitic inorganic-organic polymer electrolyte (Z-IOPE precursor) synthesized as described elsewhere.[33] Z-IOPEs consist of clusters of the desired metals crosslinked through bridges of a suitable binder.[32-33] In the preparation of *"second-generation"* ECs, the Z-IOPE impregnates the desired support; a variety of electrically-conductive systems were explored, including both nanoporous carbons [21a] and carbon black nanoparticles.[34] The precursor then undergoes a multi-step pyrolysis process, followed by suitable treatments (*e.g.*, washing in $H_2O$, or etching in HF followed by a second pyrolysis [18b]) that are crucial to modulate the physicochemical properties and the electrochemical performance of the final EC. It was demonstrated that the proposed protocol yields ECs exhibiting an outstanding performance, both in *"ex-situ"* experiments [34-35] and at the cathode of a single PEMFC in operating conditions.[21a, 36] In the *"second-generation"* ECs, the support *"core"*: (i) ensures a good electron transport, minimizing the ohmic drops; and (ii) allows for a good dispersion of the active sites, maximizing their accessibility to reactants and products.[32, 35] The CN *"shell"* plays a different role, stabilizing the active sites in C and N *"coordination nests"*. Most of the N atoms of the CN *"shell"* are actually located in the *"coordination nests"*, and stabilize the active sites without compromising the electrical conductivity of the system.[18b]

The *"core-shell"* CN-based ECs proposed in this work: (i) include graphene as the support *"core"*; and (ii) are a first attempt to implement graphene into ECs obtained with the proposed preparation protocol. This report is meant to pave the way for further studies aimed at the development of ECs capable to exploit as much as possible the unique properties of graphene to achieve improved performance and durability. The proposed ECs include two metals: (i) either Co or Ni as the *"active"* metal; and (ii) Sn as the *"co-catalyst"*.[5] Co and Ni are selected in the framework of a systematic effort to study the impact of the *"active"* metal on the morphology and ORR performance of *"Pt-free"* ECs, exploring alternatives to Fe that has attracted most of the research efforts reported in the literature.[7, 14, 37] Sn is chosen as the *"co-catalyst"* since it is likely well-stabilized in the CN *"shell"* (indeed, Sn is able to form strong and stable bonds with carbon [38]), and improves the hydrophilicity of the EC surface by forming stable oxides. The latter are expected to promote the ORR kinetics, particularly in an alkaline environment, by facilitating the first *"outer-shell"* adsorption of $O_2$ on the active sites.[39]

## 2. Results and discussion

### 2.1. Synthesis observations

The proposed ECs are obtained by a synthetic route including three main steps: (i) preparation of a precursor; (ii) multi-step pyrolysis process; and (iii) post-pyrolysis treatment and activation.[5] The precursor is prepared by impregnating the graphene nanoplatelet support Gr with a zeolitic inorganic-organic polymer electrolyte (Z-IOPE) comprising:[33] (i) anionic complexes based on dimethyltin dichloride and either $K_3Co(CN)_6$ or $K_2Ni(CN)_4$, that are bridged by (ii) sucrose as the organic binder.[32] The defects of Gr are covered mostly by hydroxyl groups (see Figure S1 in ESI), that are expected to take part in the equilibria leading to the formation of the Z-IOPE.[32] This will facilitate a homogeneous covering of the Gr support by the Z-IOPE. The multi-step pyrolysis process triggers the graphitization of the Z-IOPE by the expulsion of H, O and N heteroatoms; metal species are also nucleated in the Z-IOPE.[5] The end product is a nanocomposite material exhibiting a *"core-shell"* morphology, where the Gr support is the *"core"*, that is covered by a carbon nitride (CN) *"shell"* matrix embedding metal species in *"coordination nests"* based on C and N ligands. Indeed, N atoms are introduced exclusively through the -C≡N ligands of the $Co(CN)_6^{3-}$ or $Ni(CN)_4^{2-}$ complexes. Accordingly, it is expected that N atoms are mostly located in close proximity to the metal species nucleated in the CN *"shell"*, *i.e.*, in the *"coordination nests"*. The M1 of M1Sn$_x$-NC systems (M1 = Co, Ni in *"CoSn$_{0.5}$"* and *"NiSn$_{0.5}$"* ECs, respectively) at the interface between the CN *"shell"* and the metal species nucleated therein are expected to play a major role to promote the ORR.[40] The different post-pyrolysis treatments resulting in the final *"pristine"* and *"activated"* ECs are meant to promote the accessibility of $O_2$ to the M1Sn$_x$-NC systems, removing contaminants and *"non-active"* metal species [18b, 37] with the aim to maximize the ORR performance.

## 2.2. Chemical composition

The bulk chemical composition of the ECs is determined by inductively-coupled plasma atomic emission spectroscopy (ICP-AES) and microanalysis; the surface chemical composition is evaluated by x-ray photoelectron spectroscopy (XPS). The results are reported in Table 1 and Table 2, respectively. A comparison of the molar ratios $n_H/n_C$, $n_O/n_C$, $n_N/n_C$, $n_{M1}/n_C$, $n_{M1}/n_N$, $n_{Sn}/n_{M1}$, with M1 = Co, Ni of the proposed ECs are shown in Figure 1.

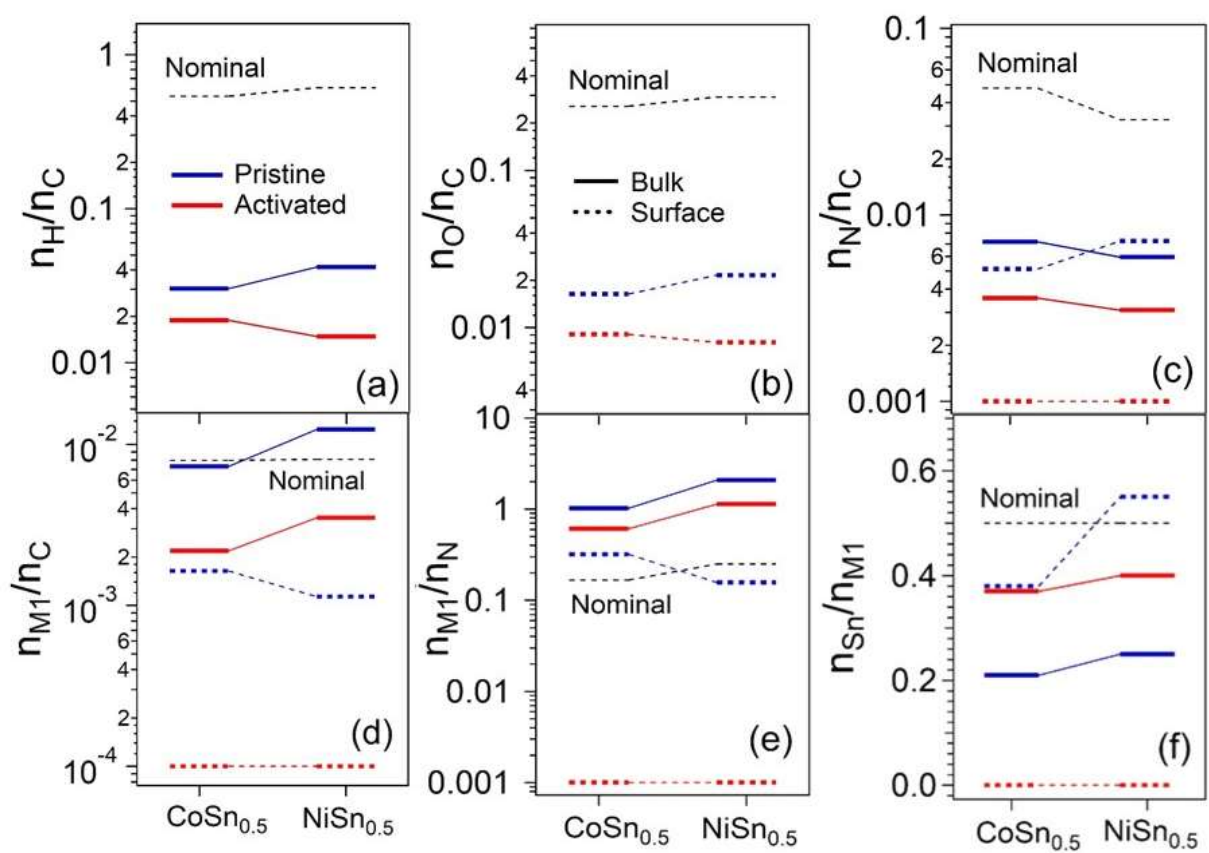

**Figure 1.** Bulk and surface chemical composition of M1Sn$_{0.5}$ CN Gr-supported ECs. The following molar ratios are displayed: $n_H/n_C$ (a); $n_O/n_C$ (b); $n_N/n_C$ (c); $n_{M1}/n_C$ (d); $n_{M1}/n_N$ (e); $n_{Sn}/n_{M1}$ (f); M1 = Co, Ni. *"Nominal"* ratios are evaluated on the basis of the stoichiometry of the reagents.

**Table 1.** Bulk chemical composition of the graphene-based ECs and the supports.

| Electrocatalyst | Weight% | | | | | | | Formula |
|---|---|---|---|---|---|---|---|---|
| | $K^{(a)}$ | $Co^{(a)}$ | $Ni^{(a)}$ | $Sn^{(a)}$ | $C^{(b)}$ | $H^{(b)}$ | $N^{(b)}$ | |
| **CoSn$_{0.5}$-CN$_l$ 900/Gr** | 0.03 | 3.11 | - | 1.32 | 86.6 | 0.22 | 0.72 | $K_{0.02}[CoSn_{0.21}C_{136.8}H_{4.14}N_{0.98}]$ |
| **CoSn$_{0.5}$-CN$_l$ 900/Gr$_A$** | 0.03 | 0.95 | - | 0.7 | 88.5 | 0.14 | 0.37 | $K_{0.04}[CoSn_{0.37}C_{457.7}H_{8.62}N_{1.64}]$ |
| **NiSn$_{0.5}$-CN$_l$ 900/Gr** | 0.10 | - | 4.84 | 2.45 | 79.9 | 0.28 | 0.56 | $K_{0.03}[NiSn_{0.25}C_{80.7}H_{3.37}N_{0.48}]$ |
| **NiSn$_{0.5}$-CN$_l$ 900/Gr$_A$** | - | - | 1.52 | 1.22 | 88.6 | 0.11 | 0.32 | $NiSn_{0.40}C_{285}H_{4.22}N_{0.88}$ |
| **Gr** | - | - | - | - | 98.2 | - | - | $C_{285}$ |
| **Pristine Gr** | - | - | - | - | 99.6 | 0.3 | - | $C_{285}H_{10.2}$ |

(a) Determined by ICP-AES
(b) Determined by microanalysis

**Table 2.** Surface chemical composition of the graphene-based ECs and the supports.

| Electrocatalyst | Atomic% | | | | | | Formula |
|---|---|---|---|---|---|---|---|
| | Co | Ni | Sn | C | O | N | |
| **CoSn$_{0.5}$-CN$_l$ 900/Gr** | 0.16 | - | 0.06 | 97.7 | 1.6 | 0.5 | $CoSn_{0.38}C_{610}O_{10}N_{3.1}$ |
| **CoSn$_{0.5}$-CN$_l$ 900/Gr$_A$** | -$^{(a)}$ | - | -$^{(a)}$ | 99.1 | 0.9 | -$^{(a)}$ | $C_{610}O_{5.5}$ |
| **NiSn$_{0.5}$-CN$_l$ 900/Gr** | - | 0.11 | 0.06 | 97.0 | 2.1 | 0.7 | $NiSn_{0.55}C_{880}O_{19}N_{6.4}$ |
| **NiSn$_{0.5}$-CN$_l$ 900/Gr$_A$** | - | -$^{(a)}$ | -$^{(a)}$ | 99.2 | 0.8 | -$^{(a)}$ | $C_{880}O_{7.1}$ |
| **Gr** | - | - | - | 97.0 | 3.0 | - | $C_{880}O_{27.2}$ |
| **Pristine Gr** | - | - | - | 96.5 | 3.5 | - | $C_{880}O_{31.9}$ |

(a) This value is lower than the detection limit of the XPS instrumentation.

The molar ratios $n_H/n_C$, $n_O/n_C$, $n_N/n_C$ are lower by *ca.* one order of magnitude or more in comparison with the nominal values expected on the basis of the stoichiometry of the reagents. This evidence indicates that a large fraction of the H, O and N heteroatoms originally included in the precursor are removed during the multi-step pyrolysis process as the CN *"shell"* undergoes graphitization. The bulk $n_H/n_C$ and $n_N/n_C$, together with the surface $n_O/n_C$ molar ratios decrease by a further factor of 3-5 upon **A**, indicating that the graphitization process of the CN *"shell"* is further promoted by the activation process.

The bulk $n_{M1}/n_C$ of pristine ECs are comparable to the nominal values, witnessing that the proposed preparation route is able to yield products embedding a well-controlled amount of *"active"* metals. On the other hand, the bulk $n_{M1}/n_N$ of pristine ECs is *ca.* 10 times larger in comparison with the nominal values. This confirms that a significant fraction of N atoms is removed during the multi-step pyrolysis process, and indicates that in pristine ECs the degree of coordination of M1 by the N atoms of the *"coordination nests"* is reduced. Finally the bulk $n_{Sn}/n_{M1}$ of pristine ECs is *ca.* 2.5 times lower than the nominal values, highlighting that Sn species are eliminated owing to the pyrolysis process and washing procedures. **A** has a very remarkable impact on the chemical composition of the ECs. In detail, the bulk $n_{M1}/n_C$ of ECs is decreased by a factor of *ca.* 3.5 upon **A**, witnessing a significant loss of M1 atoms. On the other hand, $n_{M1}/n_N$ of ECs is only decreased by a factor of *ca.* 2; this suggests that the M1 atoms left after **A** are much better coordinated by N species, and better embedded in *"coordination nests"*. $n_{Sn}/n_{M1}$ is also raised, indicating that **A** is able to selectively etch M1, while Sn is stabilized in the ECs. Finally (see Table 1), the combined wt% of Co and Sn in activated CoSn$_{0.5}$ (*i.e.*, 1.65 wt%) is significantly lower than the combined wt% of Ni and Sn in activated NiSn$_{0.5}$ (*i.e.*, 2.74 wt%). This indicates that, with respect to Ni-based species, **A** is more effective in etching Co-based species.

In pristine samples, with respect to the bulk, the surface $n_{M1}/n_C$ and $n_{M1}/n_N$ molar ratios are lower by a factor of *ca.* 5; the reverse trend is revealed for the $n_{Sn}/n_{M1}$ ratio. These results are interpreted admitting that the M1 atoms are preferably embedded in the CN *"shell"*, or exposed on the surface of its inner pores. Accordingly, they are not easily revealed by XPS, that is sensitive only to the most external surface layer of a sample.[41] On the other hand, with respect to both M1, Sn is preferably exposed on the EC surface; accordingly, it is expected to play an important role in the ORR mechanism (see Section 2.6, below). The surface concentrations of N, M1 and Sn on the activated samples fall below the detection limit of the XPS instrumentation (see Table 2, and the survey XPS profiles reported in Figure S2). This further confirms that **A** etches effectively heteroatoms and metals from the ECs.

## 2.3. High-resolution thermogravimetric studies

The high-resolution TGA profiles under an oxidizing and an inert atmosphere are reported in Figure 2 and Figure S3, respectively. The specific parameters characterizing the thermal events shown in Figure 2 are summarized in Table 3.

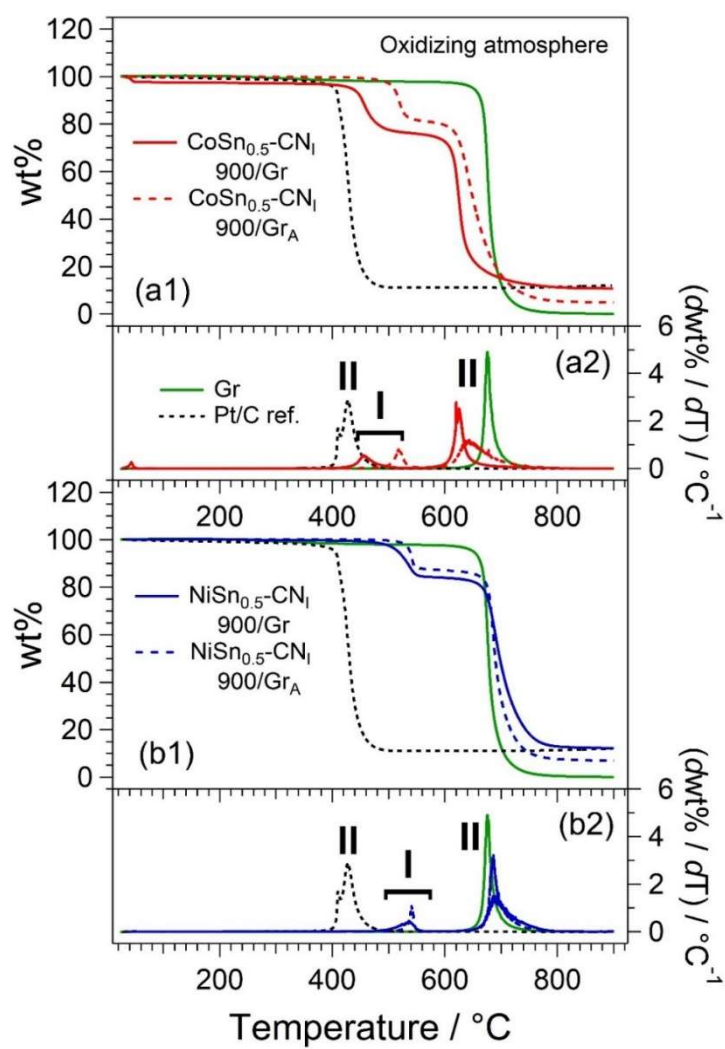

**Figure 2.** HR-TGA profiles of Gr-supported ECs in an oxidizing atmosphere: $CoSn_{0.5}$ ECs (a1) and $NiSn_{0.5}$ CN ECs (b1). The corresponding derivatives are shown in (a2) and (b2), respectively.

**Table 3.** Thermal parameters of the HR-TGA profiles in an oxidizing atmosphere measured in Figure 2.

| Sample | Event I | | Event II | | Residual / Wt% |
|---|---|---|---|---|---|
| | $T$ / °C | $\Delta Wt$ / % | $T$ / °C | $\Delta Wt$ / % | |
| **CoSn$_{0.5}$-CN$_l$ 900/Gr** | 456 | 19.1 | 624 | 67.1 | 10.8 |
| **CoSn$_{0.5}$-CN$_l$ 900/Gr$_A$** | 518 | 19.8 | 644 | 75.3 | 4.9 |
| **NiSn$_{0.5}$-CN$_l$ 900/Gr** | 533 | 16.1 | 688 | 71.7 | 12.2 |
| **NiSn$_{0.5}$-CN$_l$ 900/Gr$_A$** | 542 | 13 | 684 | 80.1 | 6.9 |
| **Pt/C ref.** | | | 427 | 88.8 | 11.2 |
| **Gr** | | | 676 | 99.5 | 0.1 |

The profiles of the proposed ECs exhibit two main events, I and II. I is associated to a mass loss between *ca.* 15 and 20 wt%, which takes place in the range 450°C < $T_I$ < 550°C and is ascribed to the decomposition of the CN *"shell"*; II corresponds to a mass loss in the 70-80 wt% range, which occurs in the temperature range 620 < $T_{II}$ < 690°C and is assigned to the degradation of the Gr support. This latter assignment is justified if we consider that the Gr support undergoes degradation at $T_{II}$ = 676°C (see Table 3). On the other hand, the event II revealed in the Pt/C ref. is attributed to the oxidative decomposition of its XC-72R support, promoted by the catalytic effect of the overlying Pt nanoparticles.[42] The high-T residue is always lower than *ca.* 12.5 wt%, and is originated by nonvolatile oxide/carbide/nitride species based on M1 and Sn that, as expected, are left after the combustion of the CN *"shell"* and of the Gr support.[43] A few general trends are clearly evident, as follows: (i) both $T_I$ and $T_{II}$ increase from CoSn$_{0.5}$ to NiSn$_{0.5}$; (ii) upon **A**, both $T_I$ and $T_{II}$ are raised while the high-T residue is decreased; and (iii) the overall mass losses associated to the events I and II are not strongly affected by M1 and **A**. The increase of $T_I$ and $T_{II}$ from M1 = Co to M1 = Ni is interpreted considering that Co is more oxophilic than Ni.[44] Accordingly, Co can better adsorb oxygen from the oxidizing atmosphere (air). This triggers the oxidative decomposition of the carbonaceous species surrounding Co at lower temperatures, yielding the thermal events I and II. It is also pointed out that this phenomenon occurs even if the wt% of Co in CoSn$_{0.5}$ ECs is lower than the wt% of Ni in NiSn$_{0.5}$ ECs (see Table 1). This suggests that Co-based species are better embedded/coordinated into the CN matrix (through the *"coordination nests"*) in comparison with the Ni-based species. Upon **A**, a significant fraction of both Co- and Ni-based species are etched from the ECs (see Table 1) and the graphitization of the ECs is improved (see Section 3.2). Accordingly, the *"activated"* ECs exhibit a much lower surface density of sites (*e.g.*, heteroatoms/defects on the CN *"shell"*, interfaces between Co-/Ni-based species and the CN *"shell"*) where the oxygen can initiate the degradation of the carbonaceous species. This results in an improved tolerance towards oxidative decomposition, that is shifted to higher temperatures. The etching of both Co- and Ni-based species upon **A** also reduces the high-T residue, that is ascribed to nonvolatile oxide/carbide/nitride species based on M1 and Sn metals (see above). The very small impact of M1 and **A** on the overall magnitude of the events I and II indicates that the proposed synthetic approach is able to control quite precisely the wt% of the CN *"shell"* of the ECs, irrespectively of the metal species characterizing the system. The analysis of the HR-TGA profiles under an inert atmosphere (see Figure S3) indicates that all the proposed ECs: (i) adsorb a negligible amount of atmospheric moisture; indeed, the low-temperature (T < 100°C) mass loss typically ascribed to this phenomenon is lower than 1 wt%; and (ii) are thermally stable up to at least 700°C. Indeed, at higher temperatures, only a very small mass loss is detected (on the order of less than 10 wt%). This latter event is ascribed to the oxidative degradation triggered by the small wt% of oxygen included in the ECs (see Table 2).

*2.4. Porosimetry/surface structure studies*

The specific areas of the pore structural features are summarized in Figure 3.

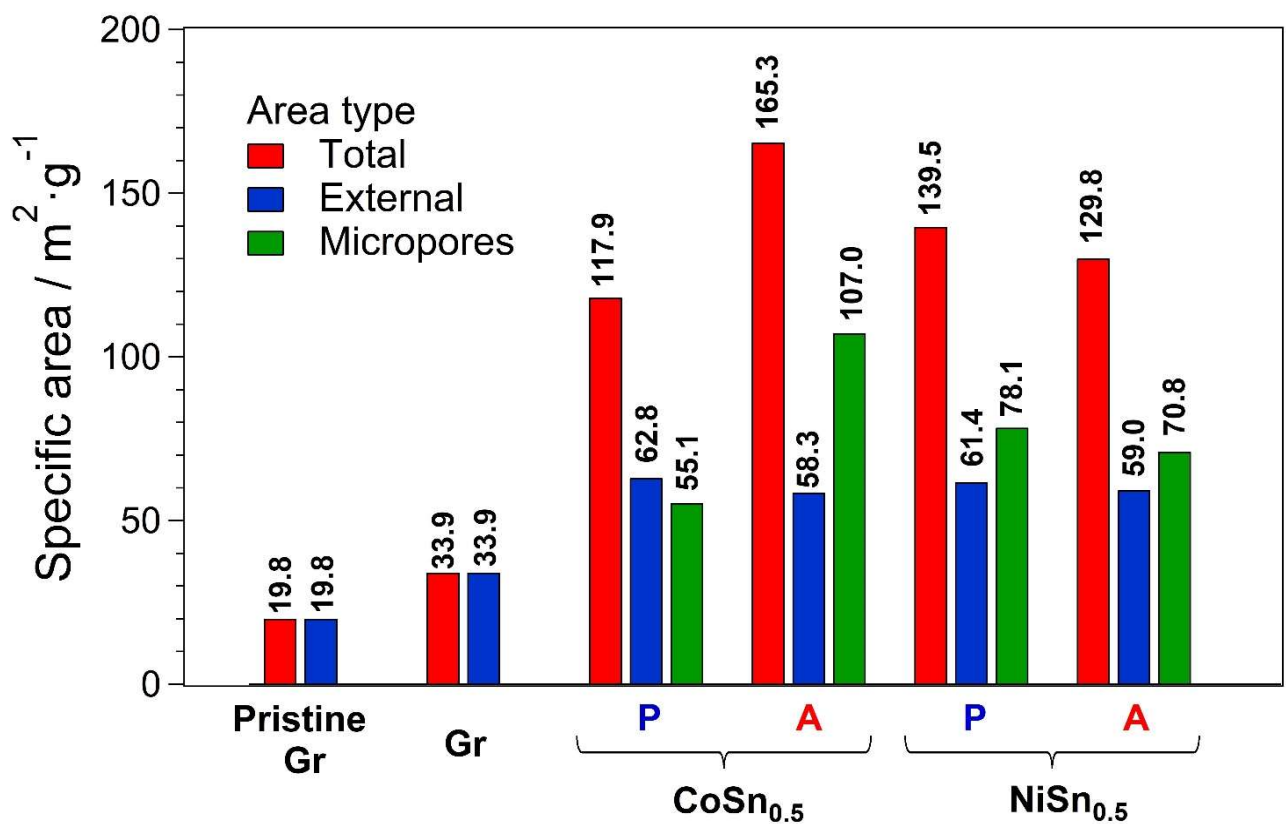

**Figure 3.** Surface area of hierarchical Gr-supported ECs as determined by nitrogen physisorption techniques. "**P**" and "**A**" are labelling the pristine and the activated ECs, respectively.

No microporosity is revealed for both the *"Pristine Gr"* and the *"Gr"* supports. This evidence is consistent with the sheet-like structure of 2D graphene.[25] The total specific area of Pristine Gr is equal to 19.8 $m^2 \cdot g^{-1}$. This value is lower in comparison with that of *"ideal"* graphene monolayer (*ca.* 2600 $m^2 \cdot g^{-1}$ [25]). This result reveals that Pristine Gr actually consists of stacked graphene layers. With respect to Pristine Gr the specific area of the Gr support, used in the subsequent synthesis of the ECs, increased by *ca.* 50% to 33.9 $m^2 \cdot g^{-1}$. This demonstrates that the synthetic procedure here adopted (see Section 2.1) is able to partially exfoliate Pristine Gr. The total specific area of the ECs is much larger in comparison with that of the Pristine Gr support, ranging from *ca.* 120 to 170 $m^2 \cdot g^{-1}$. Furthermore, all the proposed ECs exhibit some degree of microporosity. This evidence is rationalized admitting that during the preparation procedure of the ECs, a combination of the following two effects is concurring: (i) the graphene layers of the Pristine Gr support are further exfoliated; and (ii) the graphene layers of the Pristine Gr support undergo covering with a CN matrix *"shell"* that exhibits a rough morphology, and gives rise to the formation of micropores.

This latter phenomenon is also confirmed by the confocal μ-Raman spectra of the supports and ECs (see Figure S4), and in particular by the analysis of the ratio *"$I_D/I_G$"* between the intensities of the D- and the G-bands detected at *ca.* 1340 and 1580 cm$^{-1}$, respectively.[45] The $I_D/I_G$ ratio is a good indicator of the degree of disorder on the surface of a carbon-based system; it increases as the density of defects is raised.[45] The $I_D/I_G$ ratios of the Pristine Gr and of the Gr supports are low, on the order of 0.04 – 0.05, and very similar to each another (see Table S1). This result is coherent with well-ordered graphene sheets exhibiting a low surface density of defects.[45] On the other hand, the $I_D/I_G$ ratios of the ECs: (i) are much larger, showing values in the 0.17 – 0.23 range; (ii) increase from NiSn$_{0.5}$ to CoSn$_{0.5}$; and (iii) are hardly affected by **A**. These results indicate that the surface of the ECs exhibits a structure which, with respect to the Pristine Gr and of the Gr supports, is much more disordered. This is typical of porous CN matrices.[43] The increase of surface disorder from NiSn$_{0.5}$ to CoSn$_{0.5}$ ECs is a further proof that Co-based coordination species are better interacting with the CN *"shell"* matrix in comparison to the Ni-based ones. Indeed, it is expected that as a M1-based specie is coordinated by the CN *"shell"* matrix, it would act as a defect and raise the intensity of the D-band owing to the breakdown of the k-selection rule.[46] Finally, the μ-Raman spectra indicate that **A** does not influence significantly the degree of disorder of the EC surface. Thus, it is deduced that M1-based species are etched upon **A**, but this latter process does not affect the overall surface structure of the CN *"shell"* matrix.

The inspection of Figure 3 highlights a few more general trends: (i) the external area of the ECs is very similar, and is equal to *ca.* 60 m$^2$·g$^{-1}$; (ii) the micropore area of CoSn$_{0.5}$ ECs rises significantly upon **A**, from 55 to 107 $m^2 \cdot g^{-1}$; and (iii) the micropore area of NiSn$_{0.5}$ ECs is not significantly affected by **A**, remaining close to *ca.* 70 $m^2 \cdot g^{-1}$. These results are interpreted if we admit that the proposed synthetic process permits to cover the Gr supports with a CN matrix in a very reproducible way, and with a controlled overall morphology and porosity in the *"shell"*. In the case of CoSn$_{0.5}$ ECs, the large increase of the area of micropores upon **A** is interpreted on the basis of the improved strength in the coordination of Co-based species by the CN matrix in comparison with the Ni-based species. Furthermore, it is likely that **A** is able to etch a fraction of stable Co-CN coordination species, leaving behind a more microporous system. This phenomenon probably does not occur in the case of NiSn$_{0.5}$ ECs. In this case, the interactions between the Ni-based species and the CN matrix are weaker with respect to the corresponding interactions in CoSn$_{0.5}$ ECs. Thus **A**, after etching the Ni-species, does not affect significantly the morphology of the CN matrix, whose porosity does not change.

The morphology of the CoSn$_{0.5}$ and of the NiSn$_{0.5}$ ECs, both pristine and activated, is shown in Figure 4 and Figure 5, respectively. Figure 4 and Figure 5 also report the electron diffraction (ED) patterns of selected representative areas of the ECs.

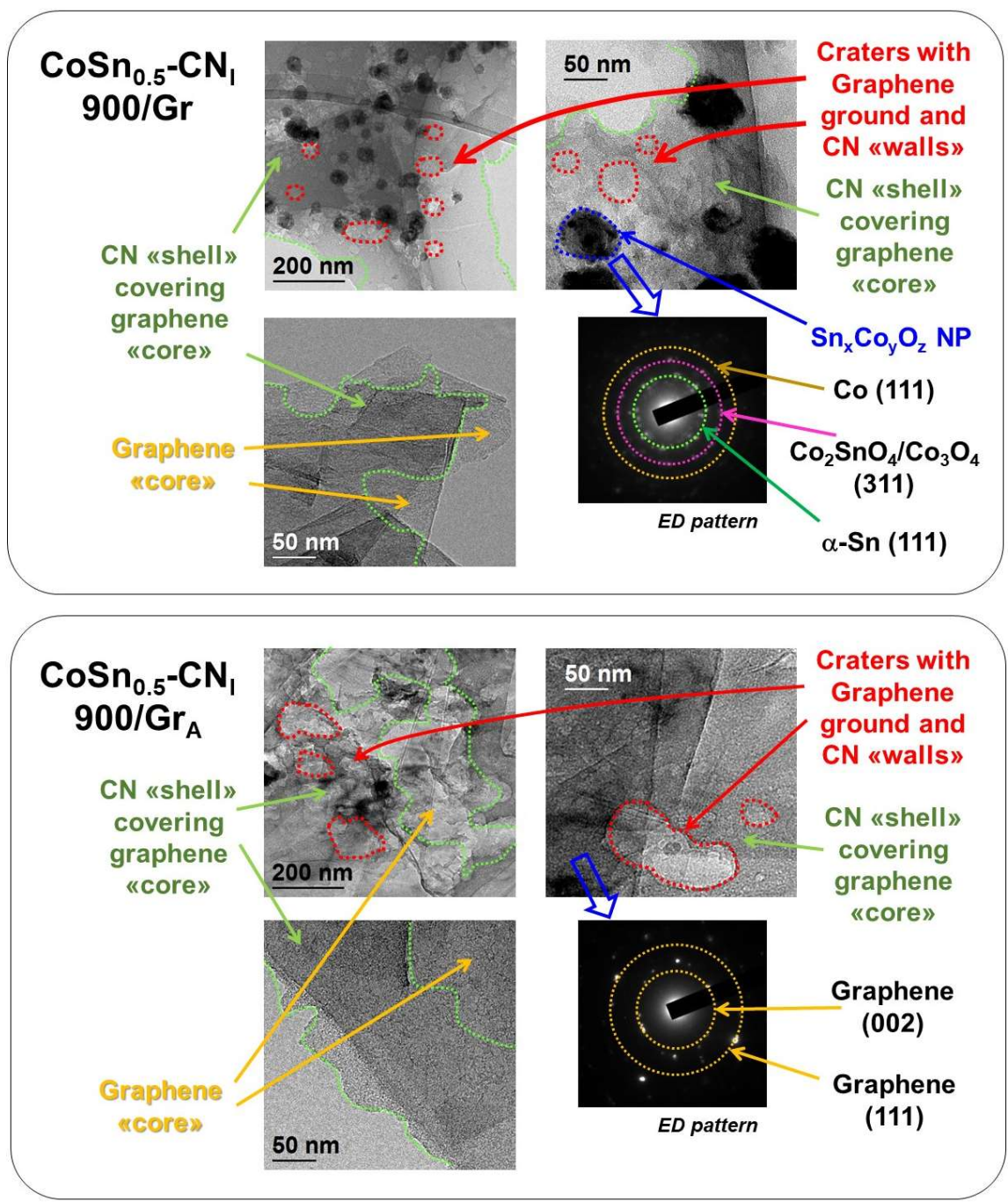

**Figure 4.** Morphology of CoSn$_{0.5}$ CN Gr-supported ECs as determined by HR-TEM. Additional structural information is revealed by selected-area electron diffraction (ED) patterns.

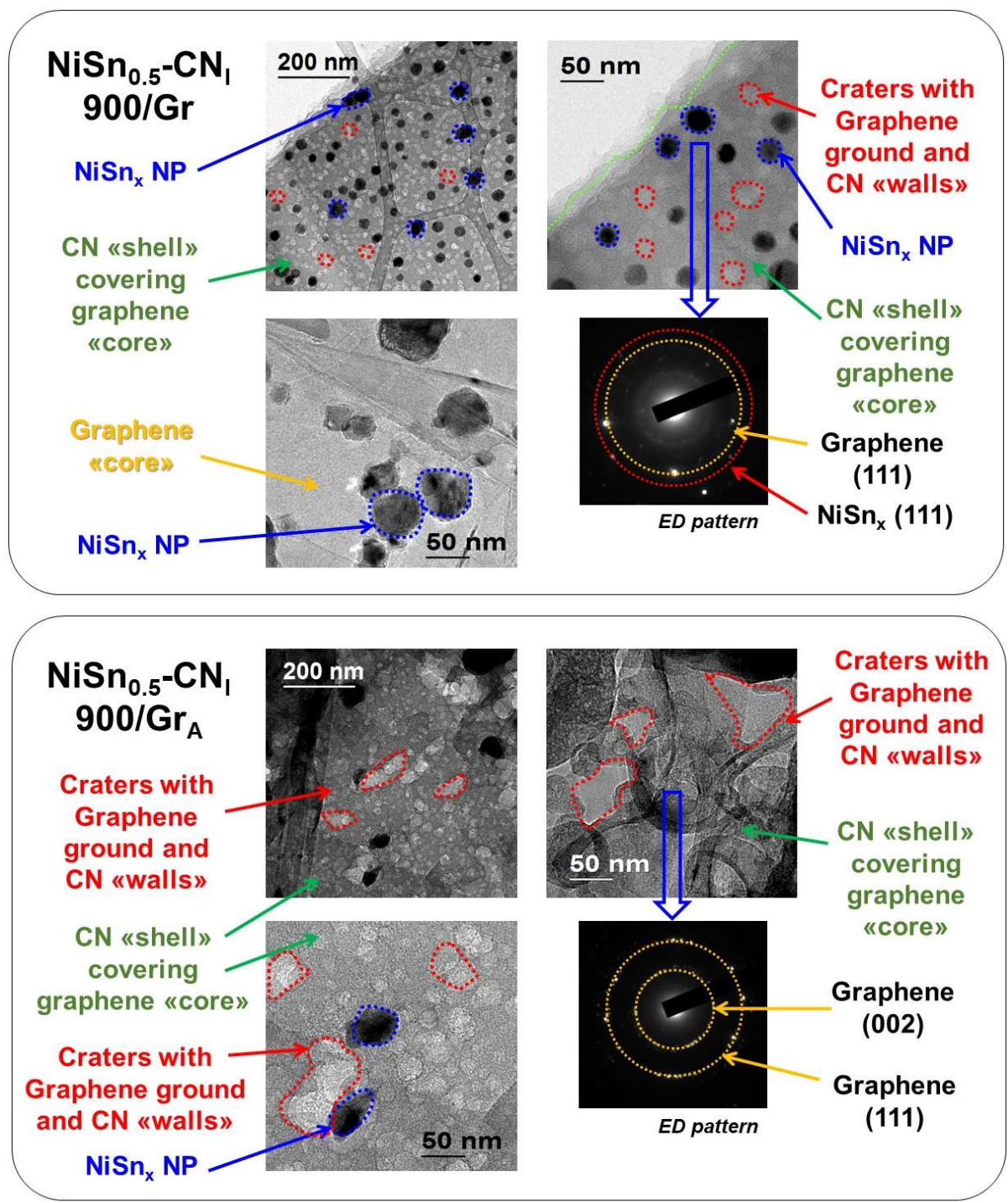

**Figure 5.** Morphology of NiSn$_{0.5}$ CN Gr-supported ECs as determined by HR-TEM. Additional structural information is revealed by selected-area electron diffraction (ED) patterns.

The ECs share several common morphology features, as follows. (i) Dark features are highlighted (d ~ 30-50 nm), which are ascribed to species based on M1 and/or Sn. (ii) A light matrix is evidenced, which is attributed to the Gr support *"core"* covered by a rough CN matrix *"shell"*. (iii) The CN *"shell"* is clearly cratered; the bottom of each crater likely consists of a graphene layer of the Gr support *"core"*, while the walls of the craters comprise the edges of the CN *"shell"*. Craters are likely obtained during the pyrolysis process, as some portions of the CN *"shell"* are flaked away. The dark features of the pristine ECs also exhibit the following differences: (i) in $CoSn_{0.5}$-$CN_l$ 900/Gr the interface between the dark features and the light matrix is rough, while in $NiSn_{0.5}$-$CN_l$ 900/Gr the interface is very neat and homogeneous; (ii) the dark features of $CoSn_{0.5}$-$CN_l$ 900/Gr are polyphasic: the ED patterns reveal features that are qualitatively ascribed to Co (*S.G.* $P6_3$/mmc, COD#9008492),[47] α-Sn (*S.G.* Fd-3m:1, COD#9008568) [47] and mixed oxides (*S.G.* Fd-3m (227), COD#9005890);[47] and (iii) the dark features of $NiSn_{0.5}$-$CN_l$ 900/Gr are probably monophasic, as the ED patterns are qualitatively attributed to a single cubic $NiSn_x$ alloy (*S.G.* Fm-3m, COD#9008476).[47] **A** affects very significantly the morphology of the proposed EC. In detail, the dark features observed in the pristine ECs are mostly removed. In the case of $CoSn_{0.5}$, the removal of the dark features is almost complete. As for $NiSn_{0.5}$, the removal is not as advanced and some dark features are still detected. This evidence is consistent with the results reported in Table 1 and Table 3, that indicate that the **A** is better capable to etch Co-based species in comparison with Ni-based species. Upon **A**, some of the CN *"shell"* is also flaked away from the Gr support *"core"*. The ED patterns of activated ECs only reveal features that can be qualitatively associated to graphene; the evident (002) and (111) reflections (*S.G.* $P6_3$/mc, COD#9008569)[47] suggest that some of the graphene layers of the Gr support adopted as the EC *"core"* are still stacked. The impact of **A** on the morphology of the CN *"shell"* is also strongly affected by M1. In detail, with respect to its pristine counterpart, in $CoSn_{0.5}$-$CN_l$ 900/$Gr_A$ the craters of the CN *"shell"* are significantly enlarged, and their edges are much rougher and more *"foamy"*. On the other hand, **A** does not enlarge significantly the craters of the CN *"shell"* of $NiSn_{0.5}$-$CN_l$ 900/Gr. The craters of the CN *"shell"* of $NiSn_{0.5}$-$CN_l$ 900/$Gr_A$ are still relatively small and well-defined. This outcome is coherent with a better and more extensive coordination in the CN matrix of the Co-based species in comparison with the Ni-based ons (see also Section 2.3 and Section 2.4). The general picture is likely the following: when there is a good coordination of the M1-based species (*i.e.*, the dark features revealed by TEM) by the CN matrix, M1-CN interactions are stronger. This imposes some constrains in the structure of the CN matrix, which gives rise to a *"rough"* interface such as in the case of $CoSn_{0.5}$-$CN_l$ 900/Gr. When **A** etches such M1-based species a small fraction of the CN matrix, which is strongly interacting with M1, is partially etched as well leaving behind the *"foamy"* edges observed in $CoSn_{0.5}$-$CN_l$ 900/$Gr_A$ EC (see Figure 4). This phenomenon is responsible of an increase in micropore area (see Section 2.4). Such an increase is not revealed for $NiSn_{0.5}$-$CN_l$ 900/$Gr_A$ (see Figure 3). In this case, we can infer that the interactions between the Ni-based species and the CN *"shell"* are weaker. Indeed, as the Ni-based species are etched upon **A**, the CN *"shell"* is mostly unaffected. In this case the structure of the craters and of the pores in general does not change significantly. Accordingly, the micropore area remains the same (see Figure 3). In conclusion, results show that craters of the CN matrix, revealed by TEM, play a pivotal role in modulating the pore structure and micropore area of the ECs.

## 2.6. CV-TF-RRDE studies

The CV-TF-RRDE profiles of the proposed ECs both in acid (0.1 M HClO₄) and in alkaline (0.1 M KOH) medium are displayed in Figure 6, together with the corresponding traces of the Pt/C ref. Capacitive contributions and ohmic drops are corrected as described in the literature.[18c, 48]

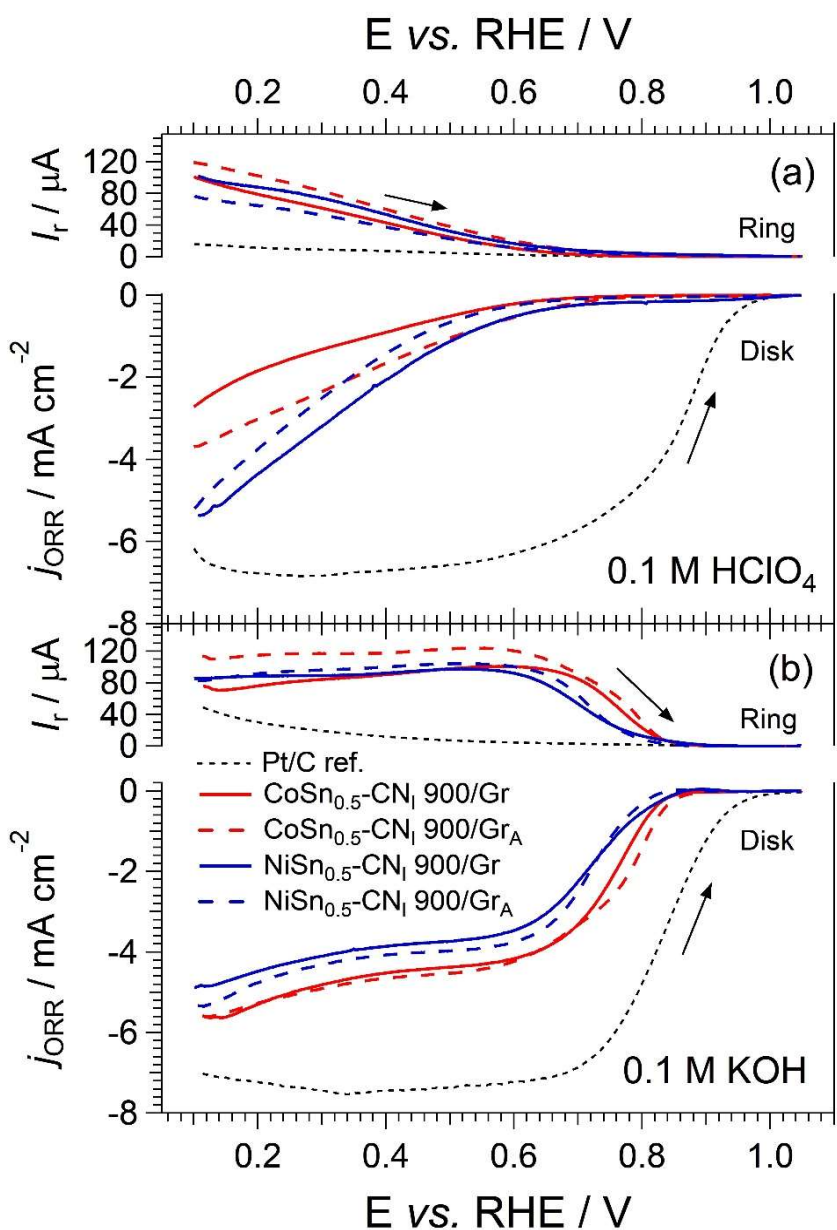

**Figure 6.** CV-TF-RRDE profiles in an O₂ atmosphere of the Gr-supported ECs. Cell filled with: 0.1 M HClO₄ (a); or 0.1 M KOH (b). T = 298 K, sweep rate = 20 mV s⁻¹, electrode rotation rate 1600 rpm, P$_{O_2}$ = 1 atm.

The geometric current density associated to the ORR detected on the RRDE disk, $j_{ORR}$, and the ring current, $I_R$, are strongly affected by the pH value. In general, the ORR overpotentials, $\eta_{ORR}$, increase in the following order: Pt/C (pH = 1; $\eta_{ORR}$ ~ 275 mV) ≤ Pt/C (pH = 13; $\eta_{ORR}$ ~ 285 mV) < ECs (pH = 13; $\eta_{ORR}$ ~ 410 mV) << ECs (pH = 1; $\eta_{ORR}$ ~ 620 mV). Starting from $j_{ORR}$ and $I_R$, the average number of electrons, n, exchanged during the ORR process is evaluated by means of Equation (1):

$$n = \frac{4I_{ORR}}{I_{ORR} + \frac{I_R}{N}} \tag{1}$$

where $I_{ORR}$ is the overall current ascribed to the ORR process as a whole. $I_{ORR}$ is obtained by multiplying $j_{ORR}$ by the geometric area of the RRDE disk ($A_{Disk} \approx 0.196$ cm$^2$). N is the collection efficiency of the RRDE ring and it is equal to 0.39. This latter value was determined experimentally in accordance with the literature.[49] The Tafel plot of the ORR is evaluated after removing the contributions ascribed to the diffusion;[18b, 49] the currents are normalized on the geometric area of the RRDE. The profiles of n *vs.* E and the Tafel plots of the ORR curves are reported in Figure 7.

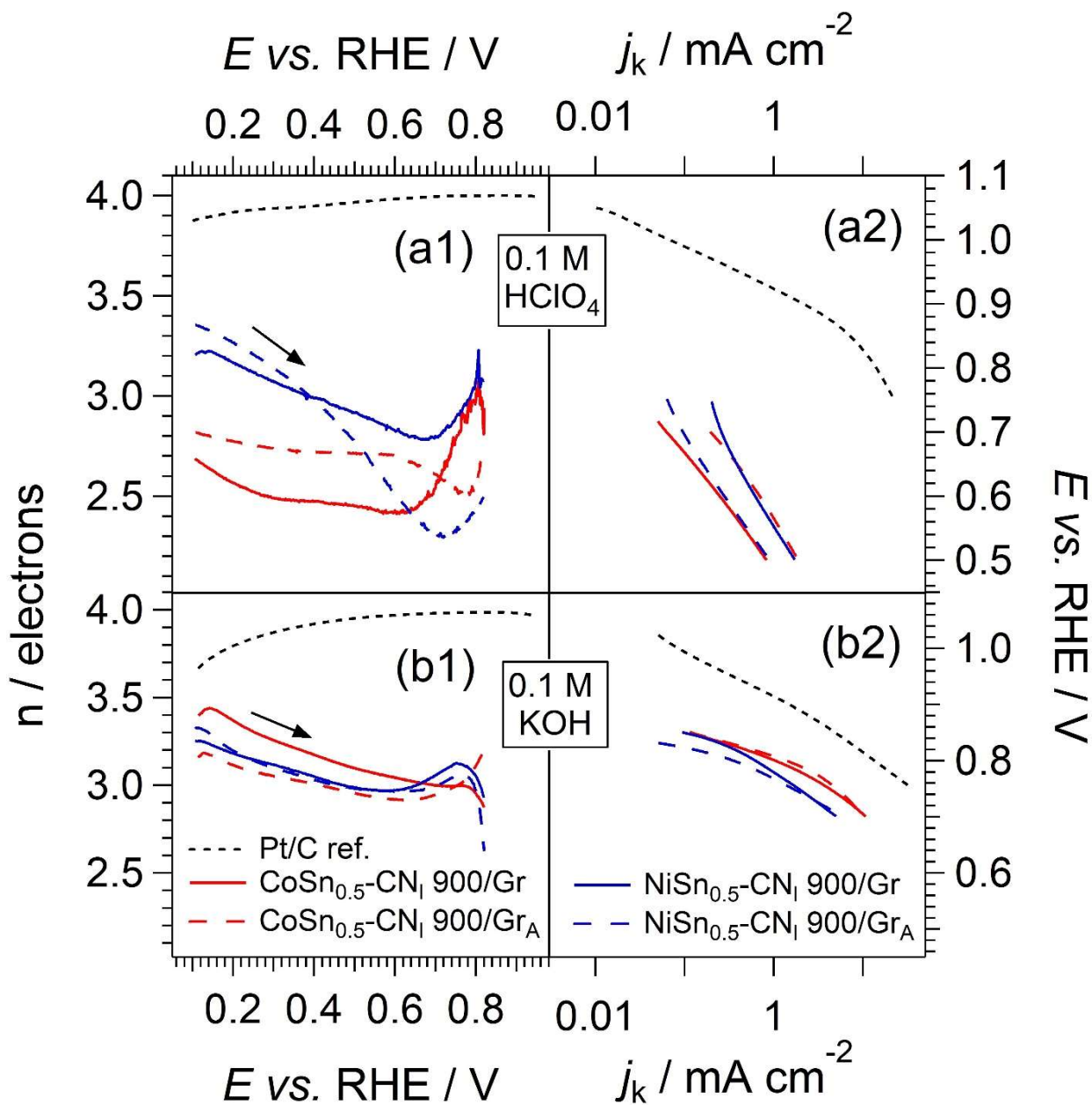

**Figure 7.** Number of electrons exchanged during the ORR (n) in a pure $O_2$ atmosphere. Cell filled with: 0.1 M $HClO_4$ (a1) and 0.1 M KOH (b1). Tafel plots of data reported in Figure 6. Cell filled with: 0.1 M $HClO_4$ (a2) and 0.1 M KOH (b2). The caption of Figure 6 reports the experimental conditions.

The figures of merit here considered to gauge the performance of the ECs in the ORR are the following: (i) number of electrons exchanged during the ORR at E = 0.3 V *vs.* RHE (n\*); and (ii) onset potential, E($j_{5\%}$), taken as the electrode potential corresponding in the ECs to a $j_{ORR}$ is equal to 5% of the maximum ORR limiting current density measured for the Pt/C ref. at *ca.* 0.3 V *vs.* RHE in the same conditions.[18b]

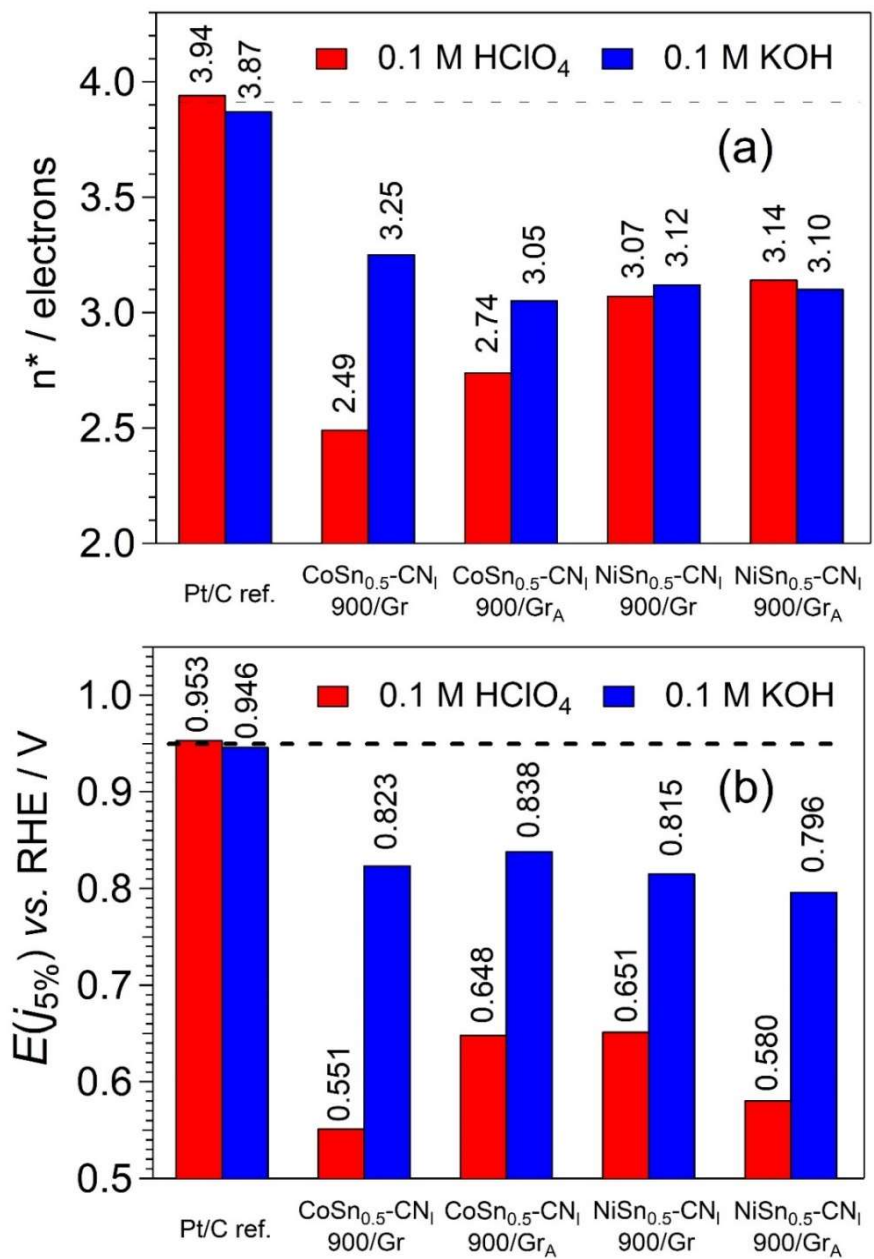

**Figure 8.** Performance of the Gr-supported ECs – figures of merit. n* – number of electrons exchanged during the ORR at E = 0.3 V *vs.* RHE (a); onset potential, E($j_{5\%}$) (b); E($j_{5\%}$) is the electrode potential corresponding in the ECs to a $j_{ORR}$ equal to 5% of the maximum ORR limiting current density measured for the Pt/C ref. at *ca.* 0.3 V *vs.* RHE in the same conditions.

The results shown in Figure 6, Figure 7, and Figure 8 are rationalized in the framework of the model adopted to describe the interplay between the physicochemical properties and the electrochemical performance in the ORR of *"Pt-free"* ECs based on a *"core"* support covered by a CN *"shell"* embedding the active sites.[18b] The overall ORR mechanism is strongly affected by the pH of the environment.[18a, 39, 50] At pH = 1, on all the proposed *"Pt-free"* ECs the ORR is bottlenecked by the first electron transfer from the EC to the dioxygen, that may occur only through an *"inner-shell"* process upon the adsorption of $O_2$ on the active sites.[18b, 39] This process is remarkably slow, and gives rise to the observed high overpotentials ($\eta_{ORR,pH = 1} \sim 620$ mV; see Figure 6(a)). Indeed, the difficult adsorption of $O_2$ on the active sites of the proposed *"Pt-free"* ECs is also witnessed by the large values of all the ORR Tafel slopes (the slopes are on the order of 180 mV·decade$^{-1}$;[50-51] see Figure 7(a2);). On the other hand, at pH = 13 the first, rate-limiting electron transfer from the EC to dioxygen may also occur through an *"outer-shell"* process.[18b] The latter likely involves most of the EC surface and is much more facile in comparison with the *"inner-shell"* process taking place at pH = 1. Correspondingly, it gives rise to a much lower overpotential ($\eta_{ORR,pH = 13} \sim 410$ mV; see Figure 6(b)). This assumption is further corroborated by the low Tafel slopes revealed at pH = 13, that fall between *ca.* 50-60 mV·decade$^{-1}$ (lower $\eta_{ORR}$, E $\geq$ 0.81 V vs. RHE) and *ca.* 120 mV·decade$^{-1}$ (higher $\eta_{ORR}$, E $\approx$ 0.75 V *vs.* RHE). This behavior is attributed to a progressive change in the $O_2$ adsorption isotherm as $\eta_{ORR}$ is raised. At low $\eta_{ORR}$, $O_2$ adsorption takes place in accordance with the Temkin isotherm (corresponding to a Tafel slope of 60 mV·decade$^{-1}$).[18c, 52] As $\eta_{ORR}$ is raised, the $O_2$ adsorption isotherm progressively changes up to the Langmuir model (corresponding to a Tafel slope of 120 mV·decade$^{-1}$).[18c, 52] It is highlighted that, with respect to the proposed *"Pt-free"* ECs, the ORR mechanism taking place on the Pt/C ref. is very different.[39] Indeed, on Pt-based active sites the ORR occurs upon a one-electron transfer, that takes place through an *"inner-shell"* process irrespectively of the pH value.[18a, 53] On Pt active sites, such a process is very facile and yields very low $\eta_{ORR}$ values, of the order of $\approx 280$ mV. As pH rises from 1 to 13, $\eta_{ORR}$ is increased by *ca.* 10 mV, as the Pt active sites are more clogged with O-based adsorbates that slightly inhibit the first $O_2$ adsorption.[39] The facile $O_2$ adsorption on Pt active sites is also witnessed by the trend observed on the Tafel slopes, that match closely that observed on the proposed *"Pt-free"* ECs in an alkaline environment.[52] One way to correlate the ORR performance with the chemical composition of the ECs is to take into consideration their *"Specific Surface Activity"*, SSA.[18b] SSA is defined as the ORR kinetic current at E = 0.85 V *vs.* RHE ($i_{k,ORR@0.85V}$) normalized on the area of the active sites of the EC.[18b] In the case of the Pt/C ref., such area corresponds to the surface area of the Pt active sites, that are vastly more efficient to promote the ORR in comparison with the carbon support. The surface area of the Pt active sites on the Pt/C ref. is determined by CO stripping measurements in accordance with the literature.[54] In the case of the proposed ECs, a precise determination of the area of the active sites is particularly troublesome due to the lack of suitable well-assessed *"probes"* capable to reveal selectively ORR active sites based on Co and Ni. However, as the ORR is carried out in the alkaline medium on *"Pt-free"* ECs, this shortcoming is mitigated since most of the EC surface is likely involved in the ORR.[18b] This is typically the case when the ORR is bottlenecked by an electron transfer occurring through an *"outer shell"* process.[39] On these bases, SSA for the alkaline environment can be evaluated by normalizing $i_{k,ORR@0.85V}$ on the entire surface of the EC as determined by nitrogen physisorption techniques; the results are reported in Table 4.

**Table 4.** SSA of the *"Pt-free"* ECs in the alkaline environment

| Electrocatalyst | SSA / $\mu A \cdot cm^{-2}$ |
|---|---|
| CoSn$_{0.5}$-CN$_l$ 900/Gr | 0.130 |
| CoSn$_{0.5}$-CN$_l$ 900/Gr$_A$ | 0.194 |
| NiSn$_{0.5}$-CN$_l$ 900/Gr | 0.090 |
| NiSn$_{0.5}$-CN$_l$ 900/Gr$_A$ | 0.023 |
| FeFe$_2$-CN$_l$ 900/C$_A$ | 1.41 [18b] |
| Pt/C ref. | 473[a] |

[a] Value referring to the Pt nanocrystals supported on the Pt/C ref; additional information reported in the literature.[18b]

With respect to the proposed *"Pt-free"* ECs, the SSA of the Pt/C ref. is significantly larger. This is consistent with a much more facile adsorption of $O_2$ on Pt-based active sites (see above). In the case of *"Pt-free"* ECs, SSA is mostly modulated by M1; in detail, SSA is raised as M1 changes from Ni to Co. This evidence is rationalized considering the higher oxophilicity of Co in comparison with Ni,[44] that facilitates the rate-determining first adsorption of dioxygen on the EC surface.[52] This interpretation is also supported by the SSA value obtained from the literature (equal to 1.41 $\mu A \cdot cm^{-2}$) referring to a relatively similar *"Pt-free"* EC including Fe-based active sites stabilized in C and N *"coordination nests"*.[18b] This SSA value, that is almost one order of magnitude larger or more in comparison with those of the *"Pt-free"* ECs discussed here, is consistent with a higher oxophilicity of Fe in comparison with both Co and Ni.[44b, 55] The data shown in this report do not allow for a clear and unambiguous identification of the role played by the Sn *"co-catalyst"* in the ORR mechanism of the ECs presented here. However, it is expected that Sn (which is present on the surface of the pristine ECs as $SnO_x$ species, see Figure S5) raises the hydrophilicity of the EC. Accordingly, $\eta_{ORR}$ is lowered since the first *"outer-shell"* electron transfer in an alkaline environment (that involves a dioxygen molecule centered on a cage of water molecules [39]) is facilitated. **A** has an important impact on SSA, that allows for an improved understanding of the correlation between: (i) the physicochemical properties of the ECs, with a particular reference to the chemical composition of the active sites; and (ii) the ORR performance. The SSA of $CoSn_{0.5}$-$CN_l$ 900/$Gr_A$ is improved in comparison with its pristine counterpart, increasing from 0.130 to 0.194 $\mu A \cdot cm^{-2}$. The reverse trend is observed for Ni-based ECs (see Table 4). This evidence is rationalized admitting that **A** etches mostly *"ORR-inert"* Co-based species that: (i) do not contribute significantly to the ORR in pristine $CoSn_{0.5}$-$CN_l$ 900/Gr; and (ii) are probably located on the surface of the dark features observed in Figure 4. On the other hand, **A** does not affect Co-based species stabilized in C and N *"coordination nests"* of the CN matrix, that thus bestow most of the ORR performance to both $CoSn_{0.5}$-$CN_l$ 900/Gr and $CoSn_{0.5}$-$CN_l$ 900/$Gr_A$. On these bases, it can be hypothesized that: (i) the Co-species that are the most active in the ORR actually consist of extremely tiny clusters of Sn and Co, stabilized in C and N *"coordination nests"*, that are located at the interface between the dark features and the CN *"shell"*; these clusters are so small (d < 1-2 nm) that are not detected by TEM (see the lower panel of Figure 4), while Co is revealed in the bulk chemical composition (see Table 1) and (ii) **A** actually improves the intrinsic ORR performance of these *"ORR-active"* Co-based species. Indeed SSA is raised upon **A**, possibly owing to the removal of impurities clogging the active sites that are left on the inner *"foamy"* walls of the craters detected in the CN matrix (see the lower panel of Figure 4). In the case of $NiSn_{0.5}$ ECs, results indicate that the Ni-based species found on the surface of the dark features revealed in Figure 5 probably provide an important contribution to the overall ORR performance of pristine $NiSn_{0.5}$-$CN_l$ 900/Gr. On the other hand any Ni-based species found at the interface between the dark features and the CN matrix, even if it survived **A**, is not expected to play a pivotal role in the ORR. This is rationalized considering that: (i) Ni-based species are not as strongly coordinated by the CN matrix as the corresponding Co-based species (see Section 2.3, Section 2.4, Section 2.5); and (ii) SSA is decreased from 0.090 to 0.023 $\mu A \cdot cm^{-2}$ (see Figure 5) as **A** etches most of the dark features. $E(j_{5\%})$ follows the same trends both in the acid and in the alkaline medium (see Figure 8(b)). Accordingly, it can be envisaged that the general picture describing the correlation between the physicochemical properties and the ORR performance in the alkaline medium is still valid in the acid medium. In conclusion, the best ORR performance in this family of *"Pt-free"* ECs is achieved in the presence of tiny, *"impurity-free"* clusters based on oxophilic metals, that are stabilized by the CN matrix (the *"shell"*) through strong coordination interactions by C and N-based ligands which are forming the *"coordination nests"*.

The overall picture outlined above is further confirmed by the average number of electrons exchanged during the ORR (see Figure 6 and Figure 7). The ORR mechanism on the active sites of the Pt/C ref. is significantly different in comparison with that taking place on the proposed *"Pt-free"* ECs, and is consistent with the results reported in the literature.[39] In summary, on the Pt active sites: (i) the ORR takes place almost exclusively with a *"direct"*, 4-electron mechanism irrespectively to the pH of the environment; (ii) n decreases slightly at E < 0.4 V *vs.* RHE. In these latter conditions the presence of adsorbates inhibits the simultaneous adsorption of the two atoms of an $O_2$ molecule on adjacent Pt sites, that is a prerequisite for direct reduction to $H_2O$.[18a, 18c] Instead, a fraction of the $O_2$ molecules manages to adsorb on only one Pt site, yielding $H_2O_2$ upon the exchange of only 2 electrons.[18a, 18c] In the case of the proposed *"Pt-free"* ECs, the following main overall trends are revealed: (i) n is significantly lower than 4 (it is on the order of 2.5 – 3.5); and (ii) n increases as E decreases. This behavior is rationalized admitting that in the proposed ECs the ORR takes place mostly owing to a two-step mechanism. (i) In the first step, $O_2$ is reduced to either $H_2O_2$ (pH = 1) [18a] or $HO_2^-$ (pH = 13) [56] with an exchange of 2 electrons. (ii) In the second step, both intermediates undergo a further reduction process that yields $H_2O$ upon the exchange of 2 more electrons.[18a] (iii) A direct, single-step dissociative adsorption of $O_2$ cannot be excluded. Both (ii) and (iii) are progressively promoted as the overpotentials associated to the corresponding processes are raised, *i.e.*, as E is lowered. At pH = 1, both M1 and **A** have a clear impact on n, as revealed in Figure 7. In detail, n increases: (i) as M1 changes from Co to Ni; and (ii) upon **A**. This evidence is rationalized admitting that, in an acid medium, the 2-electron transfer from the electrode to the $H_2O_2$ produced after the first reduction step takes place by means of an *"inner shell"* process.[57] The latter requires that $H_2O_2$ undergoes adsorption on the active sites, that are expected to include M1. This process is likely hindered by oxygen functionalities, that: (i) are more prevalent on Co-based than on Ni-based species (owing to the higher oxophilicity of the former element [44]); and (ii) are removed during **A** upon etching of the M1-based species (see Section 2.3, Section 2.4 and Section 2.5). On the other hand, at pH = 13, n is very similar for all the proposed ECs (see Figure 7). Accordingly, it is envisaged that in these conditions the 2-electron transfer to the $HO_2^-$ intermediate takes place through an *"outer-shell"* process. The latter is mostly insensitive to a particular adsorption site and is likely not strongly affected by the chemical composition of the surface of each EC.[18b]

## 3. Conclusions

In this work it is demonstrated that: (i) the synthetic route can be successfully adopted in the preparation of *"core-shell"*, *"Pt-free"* ECs including hierarchical graphene-based support *"cores"*; and (ii) important insights are obtained on the interplay between the physicochemical properties of *"Pt-free"* ECs and the electrochemical performance.

Graphene platelets act as the *"core"*, which is covered by a microporous and cratered CN matrix (the *"shell"*), that embeds M1-based species (M1: Co, Ni) in C and N-based ligand *"coordination nests"*. M1 acts as the *"active metal"*, while Sn is introduced as a *"co-catalyst"*, with the purpose to enhance the performance in the ORR.[5] The ECs undergo a further *"activation process"*, **A**, that strongly affects the chemical composition, morphology and electrochemical performance of the ECs. In detail, **A** etches: (i) most light heteroatoms from the CN *"shell"*; indeed, $n_H/n_C$, $n_O/n_C$ and $n_N/n_C$ after **A** decrease by a factor of 3-5; and (ii) a significant fraction of M1. The M1 atoms left after **A** are more strongly coordinated by N ligands of CN matrix; thus, they are likely strongly embedded in the *"coordination nests"*. On the other hand, Sn is stabilized by the CN matrix of the ECs. With respect to the Ni-based species, the Co-based species are interacting more strongly in the *"coordination nests"* with the ligands of the CN *"shell"*. Indeed, upon **A**, some of the CN domains directly interacting with the Co-based species are removed. Thus, larger and more *"foamy"*-edged craters are formed, which raise the BET surface area of the micropores. These phenomena are not observed in $NiSn_{0.5}$ ECs.

The performance in the ORR of the proposed *"Pt-free"* ECs is mostly modulated by the oxophilicity of M1; accordingly, it increases as M1 changes from Ni to Co. In $NiSn_{0.5}$ ECs, most of the active sites are found on the surface of the Ni-based species. After **A**, the SSA in the alkaline medium is decreased from 0.090 to 0.023 $\mu A \cdot cm^{-2}$. For CoSn$_{0.5}$ ECs, it is envisaged that the ORR performance is mostly bestowed by the Co-based species stabilized by the *"coordination nests"* of the CN matrix. Indeed, **A** etches most of the *"ORR-inert"* Co-based species, leaving behind only the *"ORR-active"* species. Accordingly, in the alkaline medium the SSA of CoSn$_{0.5}$-CN$_l$ 900/Gr and CoSn$_{0.5}$-CN$_l$ 900/GrA increases from 0.130 to 0.194 $\mu A \cdot cm^{-2}$, respectively.

The pH of the environment strongly affects the ORR mechanism of the proposed ECs. At pH = 1, the ORR mostly occurs with a 2-electron mechanism where the first electron transfer takes place via an *"inner-shell"* process requiring the direct adsorption of O$_2$ on the active sites. This latter process is very slow and gives rise to high $\eta_{ORR}$, which is on the order of 620 mV. At pH = 13, the ORR takes place owing to a (2 x 2)-electron mechanism, where the first electron transfer involves a facile *"outer-shell"* process. As a result, $\eta_{ORR}$ values on the order of 420 mV are registered. The highest-performing ECs presented in this study is CoSn$_{0.5}$-CN$_l$ 900/Gr$_A$, whose E($j_{5\%}$) in an alkaline medium is 0.838 V, only *ca.* 108 mV lower in comparison with the Pt/C ref.

Taken all together, the information presented in this work demonstrates the feasibility of implementing hierarchical Gr supports in *"core-shell"* ECs for the ORR. This opens the door for the preparation of further improved ECs, able to better exploit the unique features of this 2D material (*e.g.*, its wide surface area and outstanding conductivity), and provides new insight to tailor the synthesis of high-performing ORR active sites that do not comprise PGMs.

## 4. Experimental Section

### 4.1. Reagents

Potassium hexacyanocobaltate (III), 95% is purchased from Fluka; potassium tetracyanonickelate (II) hydrate (ACS reagent) is supplied by Sigma-Aldrich and dimethyltin dichloride, 95% is obtained from ABCR. Molecular biology grade sucrose is an Alfa Aesar reagent. The pristine graphene nanoplatelets, henceforth labelled *"Pristine Gr"*, are procured from ACS Material, LLC. Potassium hydroxide (98.4 wt%), perchloric acid (67-72%), and hydrogen peroxide (35%) are purchased from VWR International, Fluka Analytical and Merck, respectively. Hydrofluoric acid (48 wt%), nitric acid (>65%), isopropanol (> 99.8 wt%) and methyl alcohol (> 99.8 wt%) are Sigma-Aldrich products. The reference EC used in this work is the EC-10 product of ElectroChem, Inc.; its nominal Pt loading is 10 wt% and is indicated in the text as *"Pt/C ref."*. Carbocrom s.r.l. provides Vulcan® XC-72R; the latter is treated with $H_2O_2$ (10 vol.%) before use. All the chemicals do not undergo further purification and are used as received. Doubly distilled water is utilized in all the experiments.

### 4.2. Synthesis of the Gr-supported ECs

### 4.2.1. Synthesis of Gr support

2 g of Pristine Gr are suspended into 30 mL of $HNO_3$ under vigorous stirring. 30 mL of $H_2O_2$ are added dropwise into the resulting suspension, that is kept under vigorous stirring. The obtained product is stirred overnight at room temperature; the solid fraction is filtered on a Buchner funnel and extensively washed with doubly distilled water until the pH of the mother waters reaches 7. The resulting solids are dried overnight in a ventilated oven at a temperature of 150°C yielding the graphene nanoplatelet support, henceforth labelled *"Gr"*.

*4.2.2. Synthesis of CoSn$_{0.5}$ CN Gr-supported ECs*

300 mg of sucrose are dissolved into *ca.* 40 mL of methanol; 300 mg of Gr are added to the resulting solution, yielding a dispersion that is homogenized with a probe sonicator (Sonoplus Bandelin HD 2200; duty cycle 0.2, 2 minutes) and eventually transferred into a beaker made of Teflon®. The dispersion is heated to *ca.* 60°C by means of an oil bath and brought to a small volume (*ca.* 2 mL) under stirring; subsequently, the product is allowed to cool reaching room temperature. 100 mg of potassium hexacyanocobaltate (III) are dissolved into the minimum amount of water (*ca.* 0.5 mL); the resulting solution is added dropwise to the dispersion including Gr and the product is homogenized using a probe sonicator. 33 mg of dimethyltin dichloride are dissolved into the minimum amount of water (*ca.* 0.5 mL); a solution is yielded, that is finally added dropwise to the above product. The obtained dispersion undergoes the following steps: (i) extensive homogenization with a probe sonicator, as above; (ii) stirring for 24 h, followed by 24 h rest; and (iii) drying in a ventilated oven for 120°C. A solid pellet of precursor is obtained and inserted into a quartz tube, that is placed under a dynamic vacuum of *ca.* 1 mbar obtained with a rotary pump. The pellet of precursor then undergoes a three-step pyrolysis process, as follows: (i): 150°C, 7 hours; (ii): 300°C, 2 hours; (iii) 900°C, 2 hours. The product is divided into two aliquots. The former is treated three times with water and dried in a ventilated oven at 120°C, yielding the *"pristine"* EC labeled *"CoSn$_{0.5}$-CN$_l$ 900/Gr"* in accordance with the nomenclature proposed elsewhere.[5] The latter aliquot undergoes a treatment with HF 10 wt% lasting 2 h, it is thoroughly washed with water and is transferred into a quartz tube. The latter is placed under a dynamic vacuum of *ca.* 1 mbar obtained with a rotary pump and undergoes a 2 h pyrolysis process at 900°C. The product of this activation process (**A**) is the *"activated"* EC labeled *"CoSn$_{0.5}$-CN$_l$ 900/Gr$_A$"*. The two ECs CoSn$_{0.5}$-CN$_l$ 900/Gr and CoSn$_{0.5}$-CN$_l$ 900/Gr$_A$ are collectively indicated in the text as *"CoSn$_{0.5}$ CN Gr-supported ECs"* or, in short, as *"CoSn$_{0.5}$"*.

*4.2.3. Synthesis of NiSn$_{0.5}$ CN Gr-supported ECs*

The synthesis of the *"pristine"* EC labeled *"NiSn$_{0.5}$-CN$_l$ 900/Gr"* and of the corresponding *"activated"* EC labeled *"NiSn$_{0.5}$-CN$_l$ 900/Gr$_A$"* is carried out exactly as described in Section 4.2.2, above. The only difference is that 94 mg of potassium tetracyanonickelate (II) hydrate are used in the synthesis instead of potassium hexacyanocobaltate (III). The two ECs NiSn$_{0.5}$-CN$_l$ 900/Gr and NiSn$_{0.5}$-CN$_l$ 900/Gr$_A$ are collectively indicated in the text as *"NiSn$_{0.5}$ CN Gr-supported ECs"* or, in short, as *"NiSn$_{0.5}$"*.

The *"Instruments and methods"*, and the experimental details on the *"Electrochemical experiments"* are described in detail in the Electronic Supplementary Information (ESI).

**Acknowledgments**

The research leading to these results has received funding from: (a) the European Commission through the Graphene Flagship – Core 1 project [Grant number GA-696656]; and (b) the Strategic Project of the University of Padova *"From Materials for Membrane-Electrode Assemblies to Electric Energy Conversion and Storage Devices – MAESTRA"* [protocol STPD11XNRY]. V.D.N. thanks the University Carlos III of Madrid for the *"Càtedras de Excelencia UC3M-Santander"* (Chair of Excellence UC3M-Santander).

**TOC TEXT AND GRAPHICS**

New *"core-shell"*, *"Pt-free"* electrocatalysts (ECs) for the oxygen reduction reaction (ORR) are devised. The graphene *"core"* is covered by a carbon nitride *"shell"* matrix embedding active sites based on Sn and either Co or Ni. The interplay between the chemical composition, morphology, structure and the ORR performance and mechanism is elucidated.

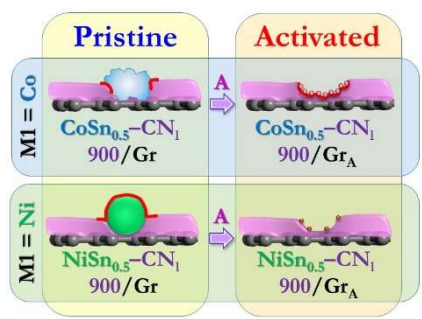